\documentclass[XXX,endnotes]{usetex-v1}
\usepackage{epsfig}     
\usepackage{url}        
\usepackage{amsmath, amsthm, amssymb}

\begin{document}

\title{Reliability Models for Highly Fault-tolerant Storage Systems}

\docstatus{Submitted to USENIX FAST Conference 2010}

\author{
\authname{Jason Resch}
\authaddr{Development}
\authaddr{Cleversafe, Inc.}
\authaddr{Chicago, Illinois, 60606}
\authurl{\url{jresch@cleversafe.com}}
\authurl{\url{http://www.cleversafe.com/}}
\and
\authname{Ilya Volvovski}
\authaddr{Cleversafe, Inc.}
\authurl{\url{ivolvovski@cleversafe.com}}
} 

\maketitle

\begin{abstract}

We found that a reliability model commonly used to estimate \emph{Mean-Time-To-Data-Loss} ($MTTDL$), while suitable for modeling RAID 0 and RAID 5, fails to accurately model systems having a fault-tolerance greater than 1.  Therefore, to model the reliability of RAID 6, Triple-Replication, or $k$-of-$n$ systems requires an alternate technique.  In this paper, we explore some alternatives, and evaluate their efficacy by comparing their predictions to simulations.  Our main result is a new formula which more accurately models storage system reliability.

\end{abstract}

\section{Introduction}

The reliability of storage systems is usually a foremost concern to implementers and users.  Data loss events can be extremely costly, consider for example the value of data in systems storing financial or medical records.  For this reason, storage systems must be engineered such that the chance of an irrecoverable data loss is extremely low, perhaps on the order of one chance in million per year of operation.  These immensely high reliabilities, however, prohibit real-world testing due to the fact that it would require a huge number of these systems to be evaluated for an extremely long time to empirically measure with any degree of accuracy.  Therefore, implementers of storage systems must rely on mathematical models for gauging the reliability of their designs.

It is important that the model used neither over- nor under-estimate reliability.  If the model overestimates reliability, then the system will be more prone to data loss than expected.  If the model underestimates reliability, the system will be designed with an excessive level of fault-tolerance and therefore be overly expensive.  We evaluated two models used in reliability analysis, one presented by Chen et al.\cite{reliability} and another given by Angus\cite{markov}. We found that while the Chen and Angus methods agree for systems with a fault-tolerance of zero or one, beyond that, the models diverge by a factorial of the system's fault-tolerance.

This paper is organized as follows:  In the \emph{Background} section, we introduce basic concepts in reliability analysis and define the meaning of notation used throughout this paper.  Next we introduce two models used in reliability analysis, one presented by Chen et al. which is commonly used in the analysis of data storage systems and a more general model presented by Angus for analyzing the reliability of $k$-of-$n$ systems.  To judge the applicability of these models we show simulations of how long systems take to fail and discuss the results.  Following that, we derive a new model which the simulation shows to have superior accuracy to both the Chen and Angus models.  Lastly, we mention some areas for further improvement to the model we give.

\section{Background}

The term \emph{reliability}, as used in this paper, refers to the probability of correct operation over a given period of time, where correct operation is defined as the absence of an irrecoverable data loss.  Gibson\cite{gibson} showed that for systems with a constant failure rate, the following exponential function may be used to estimate reliability over time $t$ given the \emph{Mean-Time-To-Data-Loss} ($MTTDL$) of the system: $e^{-t/MTTDL}$.  Knowing this function, the main difficulty in estimating reliability becomes accurately estimating the system's $MTTDL$. 

In this analysis, it is important to distinguish between a data loss and an \emph{irrecoverable} data loss.  Hard drives and tapes inevitably fail and these are instances of data loss.  However, storage systems are capable of recovering from such failures so long as the number of failures does not exceed the system's fault-tolerance.  When the number of failures exceeds a system's fault-tolerance, data can no longer be read and therefore lost data can no longer be recovered.

The rate at which individual components (hard drives, tapes, etc.) fail is denoted as $\lambda$ while the rate at which those components are repaired is $\mu$.  Alternatively, these two rates might instead be expressed as times: \emph{Mean-Time-To-Failure} ($MTTF$) and \emph{Mean-Time-To-Repair} ($MTTR$) respectively.  When $\lambda$ and $\mu$ are constant over time, meaning exponentially distributed, $MTTF = \frac{1}{\lambda}$ and $MTTR = \frac{1}{\mu}$.  Throughout this paper $MTTF$ and $MTTR$ will only be used to refer to the $MTTF$ and $MTTR$ of components, never the system.

To increase a system's $MTTDL$ the $MTTF$ of components should be as long as possible while the $MTTR$ should be as short as possible.  While implementers of storage systems can choose the components to use, once selected, they have little control over the $MTTF$ or $MTTR$.  $MTTF$ is usually defined by the manufacturer of the component and little to nothing can be done to increase it.

$MTTR$, on the other hand, may be improved to an extent.  $MTTR$ is the sum of the service time and rebuild time.  By using hot spares, the service time may be reduced to near zero and by prioritizing rebuild I/O over normal I/O requests, rebuild time may be minimized.  For storage devices, however, rebuild time has an ultimate floor defined by the I/O rate of the device.  Consider that if a 1 TB disk fails, rebuilding it requires writing 1 TB of data.  If the I/O rate of the disk is 100 MB/s it will take a minimum of 2.9 hours to recover.

Given the limited control implementers have over $MTTF$ and $MTTR$ the most important consideration for reaching a target $MTTDL$ is choosing an appropriate level of fault-tolerance for the system.  The simplest method for achieving fault-tolerance in a storage system is by using replication.  That is, create some number ($n$) of copies of the data and store each copy to a different component.  A replicated data storage system can tolerate the failure of $n-1$ components, so long as one copy remains, the failed copies can be remade.  In this respect, implementers have complete control over the fault-tolerance, and by extension, the reliability of the systems they create.

Achieving fault-tolerance by making replicated copies, however, is very inefficient.  Other more advanced methods are known for achieving fault-tolerance, such as RAID 5, RAID 6, and erasure codes.  Unlike copy-based systems which require 1 of the $n$ components to remain operational, these systems require $k$ of $n$ components to remain operational, where $k = n - f$ with $f$ being fault-tolerance.

With an accurate model for estimating $MTTDL$, implementers may engineer systems to meet the reliability requirements using a minimum level of fault-tolerance.  By minimizing fault-tolerance the storage system will be more efficient as less redundant information need be stored or calculated.  All reliability models presented in this paper are capable of estimating $MTTDL$ from the four metrics: $k$, $n$, $MTTF$ and $MTTR$.

\section{Technology Trends and Implications}

One might ask, what level of fault-tolerance is sufficient for any practical purpose?  Unfortunately, no answer remains true indefinitely.  Consider that as disk capacities have grown, their performance has not kept pace.  This has resulted in a very large increase in disk repair times.  Whereas it took 57 seconds to read an entire 40 MB disk in 1991, it takes 3.3 hours on a modern 750 GB drive\cite{toms}.  To cope with these longer repair times, systems have had to increase their level of fault-tolerance.  RAID 5 was sufficient when disks could be rebuilt in minutes; now RAID 6 is required to keep an acceptable reliability as rebuilds take hours or days.

Another factor causing the fault-tolerance of storage systems to increase is the sheer size of storage systems being built today.  The resulting decrease in reliability as storage capacity increases is linear; with all else being equal, a system storing 2 PB of data on 2,000 disks has twice the chance of experiencing data loss as a system storing 1 PB on 1,000 disks\cite{reliability}.  Storing more data inherently carries a higher risk of loss, and with storage requirements doubling every 24 months\cite{idc} the reliability of all that data is halved every 24 months.

Lastly, disk failures do not always manifest as complete operational failures.  Elerath\cite{elerath} showed that a more common path to data lass is via latent failures, caused by improper writes or degradation of the media over time.  Latent failures, more commonly known as \emph{Unrecoverable Read Errors} (UREs) result when a drive is unable to correctly read some sector.  The rate at which UREs manifest is generally reported to be between $10^{-14}$ and $10^{-15}$ per bit read.  This means that even if the URE rate remains constant, as disk sizes grow the likelihood of encountering a URE will increase.

Consider a RAID 6 (8+2) array composed of 1 TB disks.  After two disk failures all eight of the remaining disks must be read perfectly without error.  With a URE rate of $10^{-14}$, the chance of being able to read this amount of data without error is given by:
$$(1 - 10^{-14})^{8 \times 8,000,000,000,000} = 52.76\%$$

This means that about half the time, the system will encounter a URE during rebuild and therefore experience data loss.  Even though RAID 6 can supposedly recover from double disk failures, factoring in UREs one finds that half the time it cannot.  Therefore the true reliability of this RAID 6 array is only marginally better than a system with a fault-tolerance of one.

The net result is that when considering increasing disk capacities, larger storage systems, and the growing risk of UREs, one may conclude that increasing levels of fault-tolerance will be required in the future to simply maintain the same level of reliability.  Therefore, it is important that the reliability model used by storage system implementers be able to accurately model highly-fault tolerant systems. 

\section{Models}

\subsection{Chen's Model}

Chen et al. presented models for estimating the MTTDL for various RAID configurations\cite{reliability}, including RAID 0 (no parity), RAID 5 (single parity) and RAID 6 (dual parity).  As founders in the field of RAID, their model has seen wide adoption by those in the storage industry.  In the paper, zero redundancy RAID 0 systems are said to have a MTTDL equal to the MTTF of individual disks divided by the number of disks:

$$\frac{MTTF}{n}$$

They further presented models for RAID 5,

$$\frac{MTTF^{2}}{MTTR \times n \times (n-1)}$$

and RAID 6 arrays:

$$\frac{MTTF^{3}}{MTTR^{2} \times n \times (n-1) \times (n-2)}$$

A clear pattern emerges in the progression of increased fault-tolerance.  Looking at the above formulas, one sees that the MTTR term is taken to the power of the fault-tolerance ($f$) while the MTTF is taken to the power of $f+1$.  To account for the multiplication of $n * (n-1)$ ... $(n-f)$ we may use the factorial operator to find: $n! \div (k-1)!$, recalling that $k = n - f$.  Therefore we may obtain a generalization of the Chen model which works for any arbitrary $k$ and $n$:

$$\frac{MTTF^{f+1} \times (k-1)!}{MTTR^{f} \times n!}$$

It is straightforward to see that when $f = 0$, one obtains the RAID 0 formula.  If one sets $f = 1$ or $f = 2$, one derives the RAID 5 or RAID 6 formulas respectively.

\subsection{Angus's Model}

J.~E.~Angus published a paper titled ``On computing MTBF for a k-out-of-n:G repairable system''\cite{markov}.  His model is more general than those given by Chen et al. but nonetheless each may be used to estimate time to failure for data storage system.  There is, however, a difference between what the Angus and Chen models attempt to calculate.  The Chen model calculates MTTDL which if we stated more generally, is the \emph{Mean-Time-To-First-Failure} (MTTFF).  This is an important distinction because after the first failure in a data storage system, the system cannot be repaired because data is lost.  The Angus model does not assume this, and therefore allows repair from cases where more than ($n-k$) devices have failed.  This is why Angus defines the result as \emph{Mean-Time-Between-Failures} (MTBF) rather than MTTFF, his model finds the average amount of time between failures over an infinite amount of time.  In many situations, the MTTFF will be very close to the MTBF, and in those cases the Angus model may be used to accurately model MTTDL.  Later in this paper, we will explore the conditions under which this assumption is not valid.

Below is the formula Angus gave in his paper.  Note that this is as it appeared in the original notation, where he used $\lambda$ and $\mu$ instead of MTTF and MTTR:

$$\frac{1}{k\lambda {{n}\choose{k}} (\lambda/\mu)^{n-k}} \times \sum_{i=0}^{n-k} {{n}\choose{i}} (\lambda/\mu)^{i}$$

If we substitue $\lambda$ and $\mu$ with the notation used by Chen, the Angus formula becomes:

$$\frac{MTTF^{n-k+1}}{k {{n}\choose{k}} MTTR^{n-k}} \times \sum_{i=0}^{n-k} {{n}\choose{i}} \left(\frac{MTTR}{MTTF}\right)^{i}$$

Note that for cases when $MTTF \gg MTTR$, as is normally the case for disk drives, the summation component of the formula rapidly converges to zero.  Given that for most cases, MTTF will be a time in years and the MTTR a time in hours, the ratio of $\frac{MTTF}{MTTR}$ will be in the thousands for typical cases and therefore, if iterations beyond $i = 0$ are ignored, the result will only deviate by a few thousandths.  This level of accuracy is acceptable for most purposes, and therefore when $MTTF \gg MTTR$ one may simplify the Angus formula as:

$$\frac{MTTF}{k * {{n}\choose{k}}} \times \left(\frac{MTTF}{MTTR}\right)^{n-k}$$

This formula looks very similar to the one given by Chen.  Each has $MTTF^{f+1}$ in the numerator, and $MTTR^{f}$ in the denominator.  Where they differ is in their treatment of the $k$ and $n$ terms.  Chen gives:

$$\frac{(k-1)!}{n!}$$

While Angus gives:

$$\frac{1}{k \times {{n}\choose{k}} }$$

By decomposing the ${{n}\choose{k}}$ term used in the Angus model one obtains:

$$\frac{(n-k)! \times k!}{k \times n!} = \frac{(n-k)! \times (k-1)!}{n!}$$

The only difference between the simplified Angus model and the Chen model is that there is an extra term in the numerator of $(n-k)!$.  Therefore the MTTDL predicted by Chen's model will be a factorial of the fault-tolerance times less than the MTTDL predicted by Angus's model.  For RAID 0, and RAID 5 systems, $0!$ and $1!$ are both $1$, so no difference is observed between their predictions.  However, for RAID 6 systems with a fault tolerance of 2, Chen's model will yield a MTTDL one half of what Angus will give.

This difference is arguably minor, but what about for highly fault-tolerant systems that are now becoming possible via erasure codes?  It was shown in the introduction that increased levels of fault-tolerance will be required for very large storage systems.  Systems have already been developed\cite{oceanstore, tahoe, wuala} which can support much higher levels of fault-tolerance.  One such system has a standard configuration of 10-of-16.  With a fault tolerance of 6, the Chen and Angus predictions will differ by a factor of $6!$ (720) when estimating this system's MTTDL.  Why is this so, and which prediction is right?

To see why these models give different predictions, it helps to look at how the models were derived.

\begin{figure}[htbp]
\begin{centering}
\epsfig{file=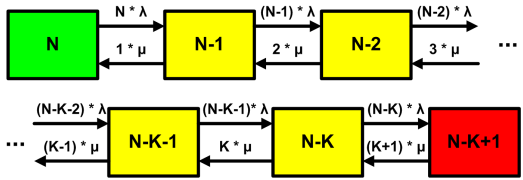, width=2.70in}
\small\itshape
\caption{\small\itshape The state transition diagram for the Angus model.  Note that failures occur at a rate equal to the number of operational devices $\times$ device failure rate and repairs occur at a rate equal to the number of failed devices $\times$ device repair rate.}
\label{fig-sample}
\end{centering}
\end{figure}

\begin{figure}[htbp]
\begin{centering}
\epsfig{file=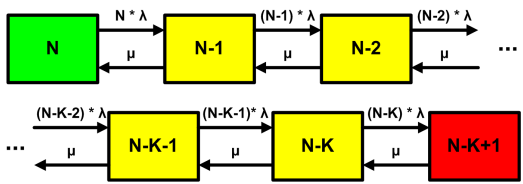, width=2.70in}
\small\itshape
\caption{\small\itshape The state transition diagram for the Chen model.  While failures occur at a rate equal to the number of operational devices $\times$ device failure rate, repairs occur at the device repair rate regardless of the number of failed devices.}
\label{fig-sample}
\end{centering}
\end{figure}

The Angus model explicitly states the assumption that there are unlimited repairmen.  This means that whether 1 device or 100 fail simultaneously, each failed device will be repaired at a constant rate.  The Chen model, on the other hand, appears to assume the per-device repair rate is inversely proportional to the number of failed disks.  There is, however, no fundamental reason why this should be the case, as each drive has its own independent I/O resources.  If two disks are simultaneously being rebuilt, the rebuild process may write to both of the disks at twice the rate it could write to a single failed disk.

Another possible consideration made in the Chen model is that because each disk has a fixed repair time, the first disk to be repaired will necessarily be the first disk to have failed.  The speed at which the first disk is repaired does not increase according to the number of failed disks, so why should the repair rate increase?  If anything is clear, it is that these models cannot both be correct.  To attempt to verify the validity of one of these two models, a Monte Carlo simulation of system failure time was created. 

\section{Simulation}

The goal of the simulation is to model the failures and repairs of $n$ independent devices until such time that more than $n-k$ devices are simultaneously in a failed state.  Failures are random and exponentially distributed over time according to the MTTF.  Repairs for each device take a constant amount of time equal to MTTR, such that MTTR time after a device's failure it will be operational.

\subsection{Pseudocode}

\begin{small}
\begin{verbatim}
random_ttf()
  return MTTF * -ln(random(0,1))

initialize()
  fail_times := fail_time[N]
  for each fail_time in fail_times:
    fail_time := random_ttf()

count_failures(start, end)
  count := 0
  for each fail_time in fail_times:
    if (start <= fail_time <= end):
      count := count + 1
  return count

simulate_time_to_data_loss()
  while True:
    nf := min(fail_times)
    count := count_failures(nf, nf+MTTR)
    if count > N-K:
      dl := time of (N-K+1)-th failure
      break
    nf := nf + MTTR + random_ttf()
  return dl
\end{verbatim}
\end{small}

The first method, \emph{random\_ttf()} generates a random failure time for a device given the device MTTF.  Because the simulation assumes failures are exponentially distributed, the negative of the natural logarithm of a random value between 0 and 1 produces a multiplier for the device MTTF.  This function is used by the \emph{initialize()} method to assign random failure times to each of the $n$ devices.  What the main simulation loop does is first check the time of the next failure, then it counts the number of failures occurring in the range from that failure time until the time that device is repaired.  If the number of failures exceeds the fault-tolerance of the system, the loop breaks, and the time of the failure that pushes the system above its fault-tolerance is returned.  Otherwise, the failure time for the device is advanced by adding the MTTR and another random time to failure.  When the loop continues, the same process will be run for the next failing device.

It is important to note that a full run of this simulation returns only one random time to data loss.  To derive an accurate estimation of the \emph{mean} time to data loss, this simulation must be run over many thousands of iterations to find the arithmetic average of all the results.  The average of the results should give a close approximation of the system's true MTTDL.

\subsection{Simulation Results}

We conducted various runs of the simulation, using different values of $n$, $k$, $MTTF$, and $MTTR$.  It was found that the magnitude of $MTTR$ was immaterial to the result of the simulation, only the ratio of $MTTF$ to $MTTR$ is important.  Therefore, for each of the results reported in the tables below, $MTTR$ is assumed to have a value of 1.  Each observed result is the average of at least 2,000 iterations of the above simulation code.  Note that for the highly fault-tolerant configurations, the MTTF had to be reduced for the simulation to complete within a reasonable period of time.

In this table, \emph{Predicted} refers to the Chen Model:

\begin{center}
  \begin{tabular}{ | r | r | r | r | r | r | }
    \hline
      N       & K        & MTTF     & Predicted    & Observed  & O/P         \\ \hline
      10      & 10       & 2000     & 2.000E2      & 1.988E2   & 0.994       \\ \hline
      10      & 9        & 2000     & 4.444E4      & 4.488E4   & 1.010       \\ \hline
      10      & 8        & 1500     & 4.688E6      & 9.446E6   & 2.015       \\ \hline
      10      & 7        & 500      & 1.240E7      & 7.786E7   & 6.278       \\ \hline
      10      & 6        & 200      & 2.511E6      & 6.407E7   & 25.513      \\ \hline
  \end{tabular}
\end{center}

Here, \emph{Predicted} refers to the Angus Model:

\begin{center}
  \begin{tabular}{ | r | r | r | r | r | r | }
    \hline
      N       & K        & MTTF     & Predicted    & Observed  & O/P         \\ \hline
      10      & 10       & 2000     & 2.000E2      & 1.988E2   & 0.994       \\ \hline
      10      & 9        & 2000     & 4.467E4      & 4.488E4   & 1.005       \\ \hline
      10      & 8        & 1500     & 9.438E6      & 9.446E6   & 1.001       \\ \hline
      10      & 7        & 500      & 7.591E7      & 7.786E7   & 1.026       \\ \hline
      10      & 6        & 200      & 6.441E7      & 6.407E7   & 0.995       \\ \hline
  \end{tabular}
\end{center}

As can be seen in the above results, Chen's model deviates from the expected roughly by a factorial of the difference between $n$ and $k$.  The Angus method, on the other hand, reasonably approximates the observed results for each case, with the simulation results never deviating more than a few percent from the prediction.

Earlier in this paper, a simplification to the Angus method was presented which can provide reasonably accurate results when the ratio between $MTTF$ and $MTTR$ is very large.  Below are the predictions of the Simplified Angus Model:

\begin{center}
  \begin{tabular}{ | r | r | r | r | r | r | }
    \hline
      N       & K        & MTTF     & Predicted    & Observed  & O/P         \\ \hline
      10      & 10       & 2000     & 2.000E2      & 1.988E2   & 0.994       \\ \hline
      10      & 9        & 2000     & 4.444E4      & 4.488E4   & 1.010       \\ \hline
      10      & 8        & 1500     & 9.375E6      & 9.446E6   & 1.008       \\ \hline
      10      & 7        & 500      & 7.440E7      & 7.786E7   & 1.046       \\ \hline
      10      & 6        & 200      & 6.027E7      & 6.407E7   & 1.063       \\ \hline
  \end{tabular}
\end{center}

Here one may observe that for cases where the MTTF is much greater than MTTR, such as with the first few configurations, the simplified Angus model produces nearly identical results to the complete Angus model.  However, for the last two cases the deviation is more pronounced, as the ratio of $\frac{MTTF}{MTTR}$ drops to a few hundred.  This deviation also manifests in the Chen model, as the prediction of the Chen model is exactly equal to the simplified Angus prediction divided by $(n - k)!$.  This explains why for the last two cases, Chen was off by slightly more than the factorial of the fault-tolerance.

\subsection{Discussion}

Given the accuracy of the Angus model compared to the Chen model, it appears the Angus model was correct to assume repair transitions in the system occur at a rate proportional to the number of failed devices.  Though the first disk to fail will be the first to be repaired and the rate of repair for this first disk does not increase as other disks fail, it is important to consider how much time is left for the repair at the time of subsequent disk failures.

Consider a RAID 6 system: after a double-disk-failure, a third will cause irrecoverable data loss.  What is the expected window of time for the system to remain in a double-disk-failure state?  The Chen model assumes that the window is MTTR, but for this to be true, both the first and second disks would had to have failed nearly simultaneously.  If we consider the average case, however, the second disk will fail when the first disk is halfway into its repair.  This is because the second failure is a random event which occurs between 0 and MTTR time after the first failure.  Therefore the true window of time for the third disk failure will not be $MTTR$, but $\frac{MTTR}{2}$.

This can explain why the Chen model underestimates MTTDL for systems with a fault-tolerance greater than 1.  This same property extends to all subsequent failures: for a system that can tolerate three failures, the third failure is expected to occur when the first disk is 2/3rds complete with its repair, making the window for the fourth failure $\frac{MTTR}{3}$.  The assumption the Angus model makes regarding increasing repair rates is therefore accurate in modeling disk failure and repair.

The expected amount of time the system spends in a state with a single failure is $MTTR$, the amount of time spent in a state with two failures is $\frac{MTTR}{2}$, and so on.  Therefore the repair rate is proportional to the number of disk failures.  Accommodating for these extra factors: 1, 2, 3 ... $(n-k)$ yields the factorial of $(n-k)!$, since each factor multiplies the $MTTDL$.

While Angus performed well for these test cases, recall that Angus does not calculate the Mean-Time-To-First-Failure, but rather the Mean-Time-Between-Failures.  It was previously mentioned that these two values are usually quite close, but this assumption only holds when there is a high probability that following recovery from a failed state the system returns to a fully operational state before again returning to the failed state.

When $MTTF \gg MTTR$ this is true, but when $MTTR$ approaches, equals or exceeds $MTTF$, the MTBF of the the system can be significantly lower than the MTTFF.  To see how Angus performs in these cases, simulations were run using much smaller $\frac{MTTF}{MTTR}$ ratios.  For these tests, a constant $n = 10$, $k = 6$ were used the results of 100,000 iterations were averaged to get the observed MTTDL.  Below are the results obtained through the Angus model:

\begin{center}
  \begin{tabular}{ | r | r | r | r | r | }
    \hline
      MTTF       & MTTR     & Predicted    & Observed  & O/P        \\ \hline
      20         & 1        & 4136.67      & 4423.75   & 1.07       \\ \hline
      10         & 1        & 205.63       & 234.28    & 1.14       \\ \hline
      1          & 1        & 0.31         & 0.67      & 2.18       \\ \hline
      1          & 10       & 0.18         & 0.65      & 3.66       \\ \hline
      1          & 20       & 0.17         & 0.65      & 3.77       \\ \hline
  \end{tabular}
\end{center}

As expected, the Angus model consistently underestimates MTTDL.  This is because the MTBF over an infinite amount of time will always be less than the MTTFF which is calculated by starting at the perfect state where all devices are operational.  When calculating the MTBF, it is not assumed that the system returns to a perfect state following failure.

To address this limitation of the Angus model in estimating MTTDL, a new model was derived with the assumption that there are no returns from the failure state, thus defining a system where the first failure is the only failure.

\section{Proposed Model}

It was found that more accurate predictions could be obtained using the following formula:

\begin{equation}
\label{mttdl}
\frac{1}{n} \sum_{i=0}^{n-k} \frac{MTTF^{i+1}}{MTTR^{i}} \sum_{j=0}^{n-k-i} \frac{{{n}\choose{j}}}{{{n-1}\choose{j+i}}}
\end{equation}

This formula and its explanation were produced and graciously offered by Yura Volvovskiy on the authors' behalf, reporting that it is a straightforward exercise in the theory of Markov chains and Poisson processes to show that the expected time to failure is given by this formula.

The Markov model used to derive this formula is identical to the one taken from the Angus paper except the rate of exit from the system failure state is zero.  In a state with $i$ disks down there are two Poisson processes.  One moves the system to the state $i+1$ when an additional disk fails.  It has intensity $(n-i) \times \lambda$.  The other brings the system to the state $i-1$ when a disk is repaired. Its intensity is $i\times\mu$.

The total intensity at any state $i$ is therefore $(n-i) \times \lambda+(i \times \mu)$.  The time spent in any given state is therefore $\frac{1}{(n-i)\times \lambda + (i \times \mu)}$.  Let us denote $T_{i}$ as the expected time to reach the failure state starting from state $i$.  $T_{i}$ will be the sum of time spent in state $i$ plus the expected time to get to the failure state from $i-1$ and $i+1$, multiplied by the respective probabilities of each transition. 

The final equation is therefore:
$$T_{i}=\frac{1}{(n-i) \times \lambda + (i \times \mu)}$$
$$+ \frac{(n-i) \times \lambda}{(n-i) \times \lambda + (i \times \mu)}\times T_{i+1}$$
$$+ \frac{(i \times \mu)}{(n-i) \times \lambda + (i \times \mu)} \times T_{i-1}$$

This provides a system of linear equations that if solved for $T_{0}$ produces the result \eqref{mttdl}.  When this model was used to generate predictions for the same configurations where the Angus model deviated, the following results were obtained:

\begin{center}
  \begin{tabular}{ | r | r | r | r | r | }
    \hline
      MTTF       & MTTR     & Predicted    & Observed  & O/P        \\ \hline
      20         & 1        & 4491.17      & 4423.75   & 0.98       \\ \hline
      10         & 1        & 246.26       & 234.28    & 0.95       \\ \hline
      1          & 1        & 0.89         & 0.67      & 0.75       \\ \hline
      1          & 10       & 0.66         & 0.65      & 0.97       \\ \hline
      1          & 20       & 0.66         & 0.65      & 0.99       \\ \hline
  \end{tabular}
\end{center}

Therefore we find that this model yields results which are much closer to those of the simulation.  In the instances where the Angus model was off by a factor of 3.77, this model was within 1\%.  One may wonder how common it is for the $\frac{MTTF}{MTTR}$ ratio to be so low.  P\^aris and Long presented a method for predicting reliability in the face of batch-correlated failures\cite{correlations}.  In it, they suggest that to model the manifestation of a batch-correlated failure, one should reduce the $MTTF$ of disks in the system to something much lower than it would be normally, suggesting a time between one week and one month might be reasonable.  In such a case, the $\frac{MTTF}{MTTR}$ ratio could be as low as 3.5.  Another reason to expect the ratio to drop is that in the past 15 years, the time it takes to read an entire hard drive has increased by over 200 times.  This means the lower bound on rebuild time, and therefore the minimum MTTR has likewise increased by this amount.  A similar decrease over the next 15 years would see the ratio drop from the thousands to around 10 to 20.

\section{Simplification and Equivalency}

With $MTTF \gg MTTR$ this new formula may be simplified by keeping only the biggest contributors to the summation.  The biggest contributor occurs for $i = n - k$.  Evaluating only this case, the formula reduces to:

$$\frac{1}{n} \times \frac{MTTF^{i+1}}{MTTR^{i}} \times \frac{{{n}\choose{0}}}{{{n-1}\choose{n-k}}}$$ 

Through some simple transformations this formula may be reweritten as:

$$\frac{1}{k} \times \frac{1}{{{n}\choose{k}}} \times \frac{MTTF^{i+1}}{MTTR^{i}}$$

Which is identical to the simplified Angus model.  Therefore when these two models are expected to produce very similar results when $MTTF \gg MTTR$, because in that case the less significant contributors to the summation converge rapidly.

\section{Conclusions}

We have demonstrated that a common model for estimating $MTTDL$ due to disk failures grossly underestimates the true $MTTDL$ for systems that have a high degree of fault-tolerance.  Furthermore, we showed that a model presented by Angus provides accurate estimations of MTTDL for systems with a high degree of fault-tolerance so long as $MTTF \gg MTTR$.

While the Angus model is more complicated than the one presented by Chen et al., a simplified version of Angus can be used which deviates by only a few percent for reasonable $\frac{MTTF}{MTTR}$ ratios. For more less constrained situations, we presented a model derived through Markov theory which exhibits a high degree of accuracy for cases of small $\frac{MTTF}{MTTR}$ ratios and showed the common relationship \eqref{mttdl} holds with the Angus model.

Our main result is that while the Chen model is adequate for systems with a fault tolerance of 0 or 1, it should not be used for systems with a fault-tolerance beyond that.  Therefore to accurately model RAID 6, Triple- (or higher) Replication, or erasure code systems, the Angus formula, or its simplified version ought to be used.  When modeling a system whose $\frac{MTTF}{MTTR}$ ratio is less than a few hundred, one should use the model \eqref{mttdl} presented in this paper over the Angus method.

\section{Future Work}

There is much room for further investigating and improvement to the method presented in this paper.  Some candidates for further research include: modeling of correlated failures, using non-exponentially distributed failures for system components, and investigating the true likelihood of Unrecoverable Read Errors in light of our findings.

\subsection{Correlated Failures}

The models presented in this paper all assume failures are statistically independent events, but much research has been done to refute this assumption in practice\cite{weibull}.  In the paper by Chen et al., a simple method for modeling correlated disk failures was presented\cite{reliability}.  What their model prescribed was to assume the MTTF for the second disk failure was 1/10th what it was for the first failure, and further assume that every subsequent disk failure is 10 times more likely than the last.

This provides reasonable results for a RAID 5 or RAID 6 system which can only tolerate one or two failures.  For RAID 6, at worst the third disk failure will only be 100 times more likely to fail than the first.  However, consider a system that could tolerate 5 disk failures.  The 6th disk failure would be modeled to have a MTTF 1/10,000th what it would be normally.  If the $\frac{MTTF}{MTTR}$ ratio is less than 10,000, adding increasing levels of fault-tolerance actually decreases the MTTDL predicted through this method.

While it is usually better to underestimate MTTDL than overestimate it, clearly this method reaches a breaking point if adding additional fault-tolerance causes the estimated MTTDL to decrease rather than increase.  To blindly follow this method's predictions, one would design a system that is less reliable than what he or she might otherwise choose.  Therefore, developing a method for modeling correlated failures in highly fault-tolerant systems would be quite beneficial.

\subsection{Exponentially Distributed Failures}

The models in this paper assume constant failure rates over time for the underlying components.  Gibson and Schroeder showed that in practice, disk failure rates follow a bathtub curve\cite{weibull} with higher levels of failure initially, stabilization during normal operating life, and slowly increasing with age.

They further found that the Weibull distribution could be used to provide a reasonable approximation of failure rates of hard drives over their useful lives.  It remains an open question how our reliability model might be amended to accommodate for hard drives with a non-exponentially distributed failure rate.

\subsection{URE Rates}

Our conjecture for why the Chen method was off for higher levels of fault-tolerance was that it fails to account for progress made in the rebuild of the first disk to have failed.  This consideration should also alter the expectation of encountering a URE.  Since on average, the first $(f-1)/f$ portion of the disks will have already been read in rebuilding the first failed disk by the time the system experiences $f$ failures.

A URE causing irrecoverable data loss, however, must happen on the part of the disks that remains to be read to rebuild the first failed disk.  Therefore, this consideration should reduce the expected likelihood of a URE causing irecoverable data loss.  Exploring exactly how the estimated MTTDL is affected remains to be explored, but we expected it to have a non-negligible effect for systems with a high degree of fault-tolerance.

\section*{Acknowledgments}

We owe a special thanks to Yura Volvovskiy who offered methodology for the calculation of \emph{Mean-Time-To-First-Failure} and solving it in general case.  This result would not be possible without his contribution.
 
We would like to take this opportunity to thank our fellow colleagues at Cleversafe for their insightful feedback and advice regarding this paper, and in particular Andrew Baptist who offered invaluable advice regarding the simulation methodology and Sanjaya Kumar for his insightful advice and feedback regarding this paper.

\end{document}